\documentstyle[12pt]{article}
\renewcommand{\baselinestretch}{1.25}

\begin{document}
\parskip=5pt plus 1pt minus 1pt

\begin{flushright}
{\bf LMU-16/95} \\
{June, 1995}
\end{flushright}

\vspace{0.4cm}
\begin{center}
{\Large\bf An Isospin Analysis of $CP$ Violation \\
in $B_{d}\rightarrow D\pi, D^{*}\pi$ and $D\rho$}
\end{center}

\vspace{0.8cm}
\begin{center}
{\bf Zhi-zhong Xing} \footnote{Electronic address:
Xing@hep.physik.uni-muenchen.de} \\
{\sl Sektion Physik, Theoretische Physik, Universit$\sl\ddot{a}$t
M$\ddot{u}$nchen, \\
Theresienstrasse 37, D-80333 M$\sl\ddot{u}$nchen, Germany}
\end{center}

\vspace{4cm}
\begin{abstract}
By use of current experimental data, we carry out an isospin analysis of the
weak decays $B\rightarrow D\pi, D^{*}\pi$ and $D\rho$. It is found that only
in $B\rightarrow D\rho$ the strong phase shift of two different isospin
amplitudes can be approximately neglected. We derive some useful relations
between the $CP$-violating measurables and the weak and strong transition
phases, and illustrate the different effects of final-state interactions
on $CP$ violation in $B_{d}\rightarrow D\pi, D^{*}\pi$ and $D\rho$.
\end{abstract}

\newpage

In the standard electroweak model, $CP$ violation is naturally described by a
non-trivial
phase in the Cabibbo-Kobayashi-Maskawa (CKM) matrix. To test the consistency of
the
standard model, the most promising way is to study the phenomena of $CP$
violation
appearing in weak decays of neutral $B$ mesons \cite{BS,REV}. Apart from a
variety of
$B_{d}$ decays to $CP$ eigenstates, some exclusive $|\Delta B|=|\Delta C|= 1$
transitions
to non-$CP$ eigenstates are also expected to have large $CP$ asymmetries
\cite{DDW,BUC}. In
most of the previous predictions of branching ratios and $CP$ violation for the
latter type
of decays, the final-state interactions were either ignored or injudiciously
simplified.
Hence an improvement of those naive and model-dependent calculations is
desirable today,
in order to yield reliable numerical results and confront them with the
experiments
in the near future.

\vspace{0.3cm}

In this note we shall follow a model-independent approach, i.e., the isospin
analysis, to
study $CP$ violation in the decay modes $B_{d}\rightarrow D\pi, D^{*}\pi$ and
$D\rho$. An
obvious advantage is that the branching ratios for some of these decays have
been measured
\cite{PDG}, and a full reconstruction of the remaining modes is available
at either existing or forthcoming $B$-meson facilities. By use of the isospin
relations
and current data, we obtain some constraints on the magnitudes of final-state
interactions
in the above decays. It is found that only in $B\rightarrow D\rho$ the strong
phase shift of
two different isospin amplitudes plays an insignificant role. We derive the
analytical
relations between the $CP$-violating measurables and the weak and strong
transition phases,
and illustrate the different effects of final-state interactions on $CP$
violation in
$B_{d}\rightarrow D\pi, D^{*}\pi$ and $D\rho$.

\vspace{0.3cm}

Let us begin with the decay modes $\bar{B}^{0}_{d}\rightarrow D^{+}\pi^{-}$,
$\bar{B}^{0}_{d}
\rightarrow D^{0}\pi^{0}$, $B^{-}_{u}\rightarrow D^{0}\pi^{-}$ and their
$CP$-conjugate
counterparts. The effective weak Hamiltonians responsible for these processes
have the isospin
structures $|1, \mp 1\rangle$. After some calculations we obtain the following
isospin relations:
$$
\begin{array}{ccccl}
M_{+-} & = & \langle D^{+}\pi^{-}|{\cal H}|\bar{B}^{0}_{d}\rangle & = &
V_{cb}V^{*}_{ud}\left (A_{3/2}
+\sqrt{2} A_{1/2}\right ) \; , \\
M_{00} & = & \langle D^{0}\pi^{0}|{\cal H}|\bar{B}^{0}_{d}\rangle & = &
V_{cb}V^{*}_{ud}\left (\sqrt{2}
A_{3/2} -A_{1/2} \right ) \; , \\
M_{0-} & = & \langle D^{0}\pi^{-}|{\cal H}|B^{-}_{u}\rangle & = &
V_{cb}V^{*}_{ud}\left (3A_{3/2}\right ) \; ;
\end{array}
\eqno(1{\rm a})
$$
and
$$
\begin{array}{ccccl}
N_{-+} & = & \langle D^{-}\pi^{+}|{\cal H}|B^{0}_{d}\rangle & = &
V_{ud}V^{*}_{cb}\left (A_{3/2}
+\sqrt{2} A_{1/2}\right ) \; , \\
N_{00} & = & \langle \bar{D}^{0} \pi^{0}|{\cal H}|B^{0}_{d}\rangle & = &
V_{ud}V^{*}_{cb}\left (\sqrt{2}
A_{3/2} -A_{1/2} \right ) \; , \\
N_{0+} & = & \langle \bar{D}^{0} \pi^{+}|{\cal H}|B^{+}_{u}\rangle & = &
V_{ud}V^{*}_{cb}\left (3A_{3/2}\right ) \; .
\end{array}
\eqno(1{\rm b})
$$
Here $A_{3/2}$ and $A_{1/2}$ correspond to the isospin $3/2$ and $1/2$
amplitudes, whose
CKM matrix elements have been factored out and whose Clebsch-Gordan
coefficients have been
absorbed into the definitions of $A_{3/2}$ and $A_{1/2}$. In obtaining eq. (1),
we have assumed
that there is no mixture of $B\rightarrow D\pi$ with other channels. It is
clear that the above
transition amplitudes form two isospin triangles in the complex plane:
$$
M_{+-} +\sqrt{2} M_{00} \; =\; M_{0-} \; , ~~~~~~~~
N_{-+} + \sqrt{2} N_{00} \; =\; N_{0+} \; .
\eqno(2)
$$
Of course $|M_{+-}|=|N_{-+}|, |M_{00}|=|N_{00}|$ and $|M_{0-}|=|N_{0+}|$ can be
directly
determined from measuring the branching ratios of $B\rightarrow D\pi$. Then we
are able to
extract the unknown quantities $A_{3/2}, A_{1/2}$ and the strong phase shift
between them
by use of eq. (1) or (2). Denoting $A_{3/2}/A_{1/2}=re^{{\rm i}\delta}$, we
relate $r$
and $\delta$ to the measurables through the following formulas:
$$
r \; =\; \frac{1}{\sqrt{3 \left (R_{+-}+R_{00}\right )-1}} \; , ~~~~~~~~
\cos \delta \; =\; \frac{r}{2\sqrt{2}}\left [ 3 \left (R_{+-}- 2R_{00}\right )
+1 \right ] \; ,
\eqno(3)
$$
where $R_{+-}=|M_{+-}/M_{0-}|^{2}$ and $R_{00}=|M_{00}/M_{0-}|^{2}$ are two
observables
independent of the uncertainty of the CKM factors. Since eq. (3) results
only from the isospin calculations, it is very useful for a model-independent
analysis of
the relevant experimental data.

\vspace{0.3cm}

Similarly we can find the same isospin relations as eqs. (1-3) for the decay
modes
$\bar{B}^{0}_{d}\rightarrow D^{*+}\pi^{-}$, $\bar{B}^{0}_{d}\rightarrow
D^{*0}\pi^{0}$,
$B^{-}_{u}\rightarrow D^{*0}\pi^{-}$; $\bar{B}^{0}_{d}\rightarrow
D^{+}\rho^{-}$,
$\bar{B}^{0}_{d}\rightarrow D^{0}\rho^{0}$, $B^{-}_{u}\rightarrow
D^{0}\rho^{-}$; and
their $CP$-conjugate processes. The existing data on the above channels
can be found in ref. \cite{PDG}. For the purpose of illustration, here we only
calculate
$R_{+-}$ and $R_{00}$ by taking the central values or upper bounds of the
measured branching
ratios. The phase space differences induced by the mass differences
$m^{~}_{D^{0}}-m^{~}_{D^{-}}$,
$m^{~}_{\pi^{0}}-m^{~}_{\pi^{-}}$ and $m^{~}_{\rho^{0}}-m^{~}_{\rho^{-}}$ are
negligible,
so is the life time difference $\tau^{~}_{B_{d}}-\tau^{~}_{B_{u}}$ \cite{PDG}.
Our results of $R_{+-}$ and $R_{00}$ are listed in Table 1. Accordingly the
lower bounds of
the isospin parameters $r$ and $\cos\delta$ can be determined, as also shown in
Table 1,
with the help of eq. (3). These results are consistent with those obtained from
the
maximum likelihood method in ref. \cite{Y}.

\vspace{0.2cm}
\begin{center}
\begin{tabular}{|l|c|c|c|c|}\hline\hline
%----------------------------------------------------------------
Decay modes $~~$	& $~~$ $R_{+-}$ $~~$	& $~~$ $R_{00}$ $~~$	& $~~$ $r$ $~~$	&
$~~$ $\cos\delta$ $~~$ \\ \hline
$~~ B\rightarrow D\pi$	& 0.566		& $<$ 0.090	& $>$ 1.02	& $>$ 0.78 \\
$~~ B\rightarrow D^{*}\pi$	& 0.500		& $<$ 0.186	& $>$ 0.97	& $>$ 0.47 \\
$~~ B\rightarrow D\rho$	& 0.582		& $<$ 0.041	& $>$ 1.07	& $>$ 0.94 \\
\hline\hline
%-----------------------------------------------------------------
\end{tabular}
\end{center}
Table 1: Estimates of the decay rate ratios ($R_{+-}$ and $R_{00}$) and the
isospin
parameters ($r$ and $\delta$) for the decay modes $B\rightarrow D\pi, D^{*}\pi$
and $D\rho$.

\vspace{0.4cm}

{}From Table 1 we observe that the magnitudes of $A_{3/2}$ and $A_{1/2}$ are
comparable in all
three cases. The current data on $B\rightarrow D\rho$ imply that
$\cos\delta^{~}_{D\rho}$ is
close to unity, i.e., there are not significant final-state interactions in
this type of decays.
In contrast, the effects of final-state interactions on $B\rightarrow D\pi$ and
in particular
on $B\rightarrow D^{*}\pi$ are non-negligible, unless the branching ratios of
$\bar{B}^{0}_{d}\rightarrow D^{0}\pi$ and $D^{*0}\pi^{0}$ are extremely
suppressed. Since
$\cos\delta \leq 1$, the lower bound of $R_{00}$ can be found from eq. (3) by
fixing the
value of $R_{+-}$:
$$
R_{00} \; \geq \; \frac{1}{2}\left (1-\sqrt{R_{+-}}\right )^{2} \; .
\eqno(4)
$$
Specifically, we have $R_{00}(D\pi)\geq 0.031$, $R_{00}(D^{*}\pi)\geq 0.043$
and $R_{00}(D\rho)
\geq 0.028$. In Fig. 1 we plot the allowed ranges of $r$ and $\cos\delta$ as
the functions of
$R_{00}$. One can see that the smaller $\delta$ is, the smaller $R_{00}$ will
be. Among the
three groups of decay modes under discussion, $B\rightarrow D\rho$ should be
the best candidate
for testing the factorization approximation and studying the $CP$ asymmetries.
Note that the lower bounds of $R_{00}$ obtained above may model-independently
isolate the
branching ratios of $\bar{B}^0_d\rightarrow D^0\pi^0, D^{*0}\pi^0$ and
$D^0\rho^0$. By use
of the central values of the available data \cite{PDG}, we get ${\rm
Br}(\bar{B}^0_d\rightarrow
D^0\pi^0)\geq 1.6\times 10^{-4}$, ${\rm Br}(\bar{B}^0_d\rightarrow
D^{*0}\pi^0)\geq 2.2\times 10^{-4}$
and ${\rm Br}(\bar{B}^0_d\rightarrow D^0\rho^0)\geq 3.8\times 10^{-4}$. It is
expected that
these three modes can soon be established in experiments.

\vspace{0.3cm}

We proceed to discuss $CP$ violation in the decay modes $B_{d}\rightarrow D\pi,
D^{*}\pi$
and $D\rho$, where every final state is common to both $B^{0}_{d}$ and
$\bar{B}^{0}_{d}$ mesons.
The $CP$ asymmetry is induced by the interplay of decay and
$B^{0}_{d}-\bar{B}^{0}_{d}$ mixing
\cite{BS,REV}. Taking $B_{d}\rightarrow D\pi$ for example, we define the
$CP$-violating
interference term via $B^{0}_{d}\rightarrow D^{+}\pi^{-}$ ($D^{0}\pi^{0}$)
versus $\bar{B}^{0}_{d}\rightarrow D^{+}\pi^{-}$ ($D^{0}\pi^{0}$) as $\xi_{+-}$
($\xi_{00}$); and that via $B^{0}_{d}\rightarrow D^{-}\pi^{+}$
($\bar{D}^{0}\pi^{0}$)
versus $\bar{B}^{0}_{d}\rightarrow D^{-}\pi^{+}$ ($\bar{D}^{0}\pi^{0}$) as
$\zeta_{-+}$
($\zeta_{00}$). Explicitly, these measurables are expressed as
$$
\xi_{+-} \; =\; {\rm Im} \left (\frac{V_{td}V^{*}_{tb}}{V_{tb}V^{*}_{td}}\cdot
\frac{M_{+-}}{\tilde{M}_{+-}}\right ) \; , ~~~~~~~~
\xi_{00} \; =\; {\rm Im} \left (\frac{V_{td}V^{*}_{tb}}{V_{tb}V^{*}_{td}}\cdot
\frac{M_{00}}{\tilde{M}_{00}}\right ) \; ,
\eqno(5{\rm a})
$$
$$
\zeta_{-+} \; =\; {\rm Im} \left
(\frac{V_{tb}V^{*}_{td}}{V_{td}V^{*}_{tb}}\cdot
\frac{N_{-+}}{\tilde{N}_{-+}}\right ) \; , ~~~~~~~~
\zeta_{00} \; =\; {\rm Im} \left
(\frac{V_{tb}V^{*}_{td}}{V_{td}V^{*}_{tb}}\cdot
\frac{N_{00}}{\tilde{N}_{00}}\right ) \; ,
\eqno(5{\rm b})
$$
where the decay amplitudes $\tilde{M}$ and $\tilde{N}$ are given by
\cite{Sachs}
$$
\begin{array}{ccccl}
\tilde{M}_{+-}  & = & \langle D^{+}\pi^{-}|{\cal H}|B^{0}_{d}\rangle & = &
V_{cd}V^{*}_{ub}\left (\tilde{A}_{3/2}-\sqrt{2}\tilde{A}_{1/2}\right ) \; , \\
\tilde{M}_{00}  & = & \langle D^{0}\pi^{0}|{\cal H}|B^{0}_{d}\rangle & = &
V_{cd}V^{*}_{ub}\left (\sqrt{2}\tilde{A}_{3/2}+\tilde{A}_{1/2}\right ) \; ;
\end{array}
\eqno(6{\rm a})
$$
and
$$
\begin{array}{ccccl}
\tilde{N}_{-+}  & = & \langle D^{-}\pi^{+}|{\cal H}|\bar{B}^{0}_{d}\rangle & =
&
V_{ub}V^{*}_{cd}\left (\tilde{A}_{3/2}-\sqrt{2}\tilde{A}_{1/2}\right ) \; , \\
\tilde{N}_{00}  & = & \langle \bar{D}^{0}\pi^{0}|{\cal
H}|\bar{B}^{0}_{d}\rangle & = &
V_{ub}V^{*}_{cd}\left (\sqrt{2}\tilde{A}_{3/2}+\tilde{A}_{1/2}\right ) \; .
\end{array}
\eqno(6{\rm b})
$$
For simplicity, we denote $\tilde{A}_{3/2}/\tilde{A}_{1/2} = \tilde{r} e^{{\rm
i}\delta}$,
where $\delta$ is the same as the strong phase shift in eqs. (1) and (3).
Subsequently we use the Wolfenstein parameters \cite{W} and the angles of the
unitarity
triangle \cite{PDG} to express the CKM matrix elements. To a good degree of
accuracy, we have $V_{ud}\approx V_{tb}\approx 1$, $V_{cd}\approx -\lambda$,
$V_{cb}\approx A\lambda^{2}$, $V_{ub}\approx
A\lambda^{3}\sqrt{\rho^{2}+\eta^{2}}
e^{-{\rm i}\gamma}$ and $V_{td}\approx A\lambda^{3}\sqrt{(1-\rho)^{2}+\eta^{2}}
e^{-{\rm i}\beta}$. Note that the parameters $(\beta, \gamma)$ are dependent
upon
$(\rho, \eta)$ through $\tan\beta =\eta/(1-\rho)$ and $\tan\gamma =\eta/\rho$.
The $CP$-violating terms $\xi_{+-}$ ($\zeta_{-+}$)
and $\xi_{00}$ ($\zeta_{00}$) turn out to be:
$$
\xi_{+-} \; =\; \frac{\left (r\tilde{r}-2\right )\sin (2\beta +\gamma) +
\sqrt{2}r
\sin [\delta -(2\beta +\gamma)] + \sqrt{2}\tilde{r} \sin [\delta +(2\beta
+\gamma)]}
{\lambda^{2}\sqrt{\rho^{2}+\eta^{2}} ~h \left
(\tilde{r}^{2}+2-2\sqrt{2}\tilde{r}\cos\delta\right )} \; ,
\eqno(7{\rm a})
$$
$$
\xi_{00} \; =\; \frac{\left (2r\tilde{r}-1\right )\sin (2\beta +\gamma) -
\sqrt{2}r
\sin [\delta -(2\beta +\gamma)] -\sqrt{2}\tilde{r} \sin [\delta +(2\beta
+\gamma)]}
{\lambda^{2}\sqrt{\rho^{2}+\eta^{2}} ~h \left
(2\tilde{r}^{2}+1+2\sqrt{2}\tilde{r}\cos\delta\right )} \; ;
\eqno(7{\rm b})
$$
and
$$
\zeta_{-+} \; =\; \frac{\left (2- r\tilde{r}\right )\sin (2\beta +\gamma) +
\sqrt{2}r
\sin [\delta +(2\beta +\gamma)] + \sqrt{2}\tilde{r} \sin [\delta
-(2\beta+\gamma)]}
{\lambda^{2}\sqrt{\rho^{2}+\eta^{2}} ~h \left
(\tilde{r}^{2}+2-2\sqrt{2}\tilde{r}\cos\delta\right )} \; ,
\eqno(8{\rm a})
$$
$$
\zeta_{00} \; =\; \frac{\left (1- 2r\tilde{r}\right )\sin (2\beta +\gamma) -
\sqrt{2}r
\sin [\delta +(2\beta +\gamma)] - \sqrt{2}\tilde{r} \sin [\delta
-(2\beta+\gamma)]}
{\lambda^{2}\sqrt{\rho^{2}+\eta^{2}} ~h \left
(2\tilde{r}^{2}+1+2\sqrt{2}\tilde{r}\cos\delta\right )} \; ,
\eqno(8{\rm b})
$$
where $h$ denotes the ratio $\tilde{A}_{1/2}/A_{1/2}$.
It should be noted that $\zeta_{-+}$ ($\zeta_{00}$) can be obtained from
$\xi_{+-}$
($\xi_{00}$) by the replacements $\beta \rightarrow -\beta$ and $\gamma
\rightarrow -\gamma$.
{}From eqs. (7) and (8) one can see that in general the strong phase shift
$\delta$ enters
the $CP$ asymmetries and plays a significant role. Only when $\delta$ is
vanishingly
small, its effect on $\xi_{+-}$ ($\zeta_{-+}$) and $\xi_{00}$ ($\zeta_{00}$)
can be safely
neglected. In this case, we have
$$
\xi_{+-} \; = \; -\zeta_{-+} \; =\; \frac{r+\sqrt{2}}{\tilde{r}-\sqrt{2}}\cdot
\frac{\sin (2\beta+\gamma)}{\lambda^{2}\sqrt{\rho^{2}+\eta^{2}} ~h} \; ,
\eqno(9{\rm a})
$$
$$
\xi_{00} \; = \; -\zeta_{00} \; =\;
\frac{\sqrt{2}r-1}{\sqrt{2}\tilde{r}+1}\cdot
\frac{\sin (2\beta+\gamma)}{\lambda^{2}\sqrt{\rho^{2}+\eta^{2}} ~h} \; .
\eqno(9{\rm b})
$$
Indeed the conditions $\xi_{+-}=-\zeta_{-+}$ and $\xi_{00}=-\zeta_{00}$ were
injudiciously
taken in most of the previous works (see, e.g., refs. \cite{DDW,BUC}).
Considering
the isospin results of $r$ and $\delta$ given in Table 1 and Fig. 1, we know
that
only $\cos\delta^{~}_{D\rho}\approx 1$ is an accetable approximation to the
current
data. Thus the $CP$ asymmetries in $B_{d}\rightarrow D\rho$ are dominated by
the
$\sin(2\beta+\gamma)$ term of $\xi_{+-}$ ($\zeta_{-+}$) or $\xi_{00}$
($\zeta_{00}$). In
contrast, the previous numerical predictions of $CP$ violation in
$B_{d}\rightarrow D\pi$ and $D^{*}\pi$ are questionable.

\vspace{0.3cm}

Now let us illustrate the effects of nonvanishing $\delta$ on $CP$ asymmetries
in
$B_{d}\rightarrow D\pi, D^{*}\pi$ and $D\rho$. Typically we take
$\lambda\approx 0.22$,
$\rho\approx -0.07$ and $\eta\approx 0.38$ \cite{AL}, and this corresponds to
$\beta\approx 19.6^{0}$ and $\gamma\approx 100.4^{0}$. Fixing $R_{+-}$, we
still use
the constraints on $R_{00}$ obtained in Table 1, eq. (4) and Fig. 1.
Since there is not any experimental information on the magnitudes of
$\tilde{A}_{3/2}$
and $\tilde{A}_{1/2}$, here we make a likely but unjustified approximation:
$h\approx 1$ and $\tilde{r}\approx r$, just for the purpose of simplicity and
illustration. The changes of
$\xi_{+-}, \zeta_{-+}$ and $\xi_{00}, \zeta_{00}$ with the allowed values of
$\cos\delta$
are shown in Figs. 2 and 3 respectively. One can see that in $B_{d}\rightarrow
D\pi$ and
$D^{*}\pi$ the effects of $\delta$ are significant, and the approximation
$\xi_{+-}=-\zeta_{-+}$ or
$\xi_{00}=-\zeta_{00}$ is invalid to a large extent. In comparison, the
final-state interactions
in $B_{d}\rightarrow D\rho$ may be negligible, if the decay rate of
$\bar{B}^{0}_{d}\rightarrow
D^{0}\rho^{0}$ is further suppressed to allow for
$\cos\delta_{D\rho}\rightarrow 1$.
Although the strong phase shift $\delta$ plays a non-negligible role in each of
the above
channels, its effect can be well isolated after the measurements of
$\bar{B}^{0}_{d}\rightarrow
D^{0}\pi^{0}, D^{*0}\pi^{0}$ and $D^{0}\rho^{0}$. Thus the determination of
$CP$ violation
in $B_{d}\rightarrow D\pi, D^{*}\pi$ and $D\rho$ remains promising in the near
future.

\vspace{0.3cm}

The experimental scenarios for observing $CP$ violation in the exclusive
$|\Delta B|
=|\Delta C|=1$ transitions have been discussed in the literature (see, e.g.,
ref.
\cite{REV}). The basic signal for $CP$ violation between a channel (e.g.,
$B^{0}_{d}\rightarrow
D^{+}\pi^{-}$) and its $CP$-conjugate counterpart ($\bar{B}^{0}_{d}\rightarrow
D^{-}\pi^{+}$)
is a nonvanishing ratio of the difference to the sum of their decay rates. For
either
coherent or incoherent $B^{0}_{d}\bar{B}^{0}_{d}$ decays to the non-$CP$
eigenstates under
study, the $CP$ asymmetries are always proportional to ($\xi_{+-}-\zeta_{-+}$)
or
($\xi_{00}-\zeta_{00}$), as shown in ref. \cite{X}. Both the time-integrated
and time-dependent
measurements are available to establish the $CP$-violating signals in
$B_{d}\rightarrow D\pi,
D^{*}\pi$ or $D\rho$, after $10^{7-8}$ $B^{0}_{d}\bar{B}^{0}_{d}$ events have
been accumulated.

\vspace{0.5cm}

I would like to thank Prof. H. Fritzsch for his constant encouragement. My
gratitude goes also
to Prof. G. Ecker and Prof. H. Pietschmann at the Institut f$\rm\ddot{u}$r
Theoretische Physik
of Universit$\rm\ddot{a}$t Wien, where this paper was written, for their warm
hospitality.
A stimulating discussion with Prof. D.D. Wu is greatly appreciated.
Finally I am indebted to the Alexander von Humboldt-Stiftung for
its financial support.

\newpage

\begin{figure}
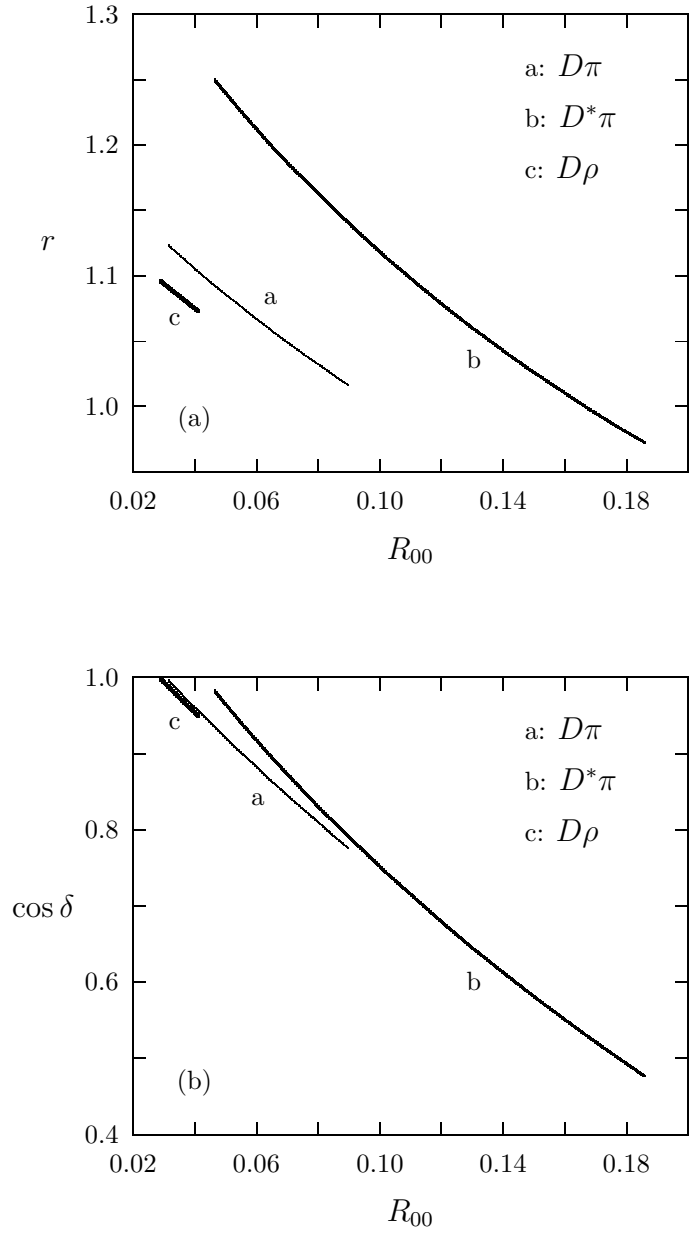

% GNUPLOT: LaTeX picture
\setlength{\unitlength}{0.240900pt}
\ifx\plotpoint\undefined\newsavebox{\plotpoint}\fi
\sbox{\plotpoint}{\rule[-0.175pt]{0.350pt}{0.350pt}}%

\vspace{0.5cm}
\caption{Possible ranges of the isospin parameters $r=|A_{3/2}/A_{1/2}|$
and $\delta=\arg (A_{3/2}/A_{1/2})$ allowed by the current data on
$B\rightarrow D\pi, D^{*}\pi$ and $D\rho$.}
\end{figure}

\begin{figure}
% GNUPLOT: LaTeX picture
\setlength{\unitlength}{0.240900pt}
\ifx\plotpoint\undefined\newsavebox{\plotpoint}\fi
\sbox{\plotpoint}{\rule[-0.175pt]{0.350pt}{0.350pt}}%
%
\vspace{0.5cm}
\caption{Illustration of the changes of the $CP$-violating terms $\xi_{+-}$ and
$\zeta_{-+}$ with
the allowed values of the strong phase shift $\delta$ in the decay modes
$B_{d}\rightarrow D\pi, D^{*}\pi$ and $D\rho$.}
\end{figure}

\begin{figure}
% GNUPLOT: LaTeX picture
\setlength{\unitlength}{0.240900pt}
\ifx\plotpoint\undefined\newsavebox{\plotpoint}\fi
\sbox{\plotpoint}{\rule[-0.175pt]{0.350pt}{0.350pt}}%
%
\vspace{0.5cm}
\caption{Illustration of the changes of the $CP$-violating terms $\xi_{00}$ and
$\zeta_{00}$ with
the allowed values of the strong phase shift $\delta$ in the decay modes
$B_{d}\rightarrow D\pi, D^{*}\pi$ and $D\rho$.}
\end{figure}

\end{document}